 \documentclass[prl,twocolumn,showpacs,twocolumngrid,superbib]{revtex4}
 \newcommand{\commentoutB}[1]{}
 \newcommand{\commentoutA}[1]{#1}
\usepackage{graphicx}
\usepackage{amsmath}
\usepackage{bm}
\usepackage{alltt}

\begin{document}

\title{\emph{Ab-inito} Linear Scaling Response Theory: \\ 
             Electric Polarizability by Perturbed Projection }

\author{Val\'ery Weber}
\altaffiliation[Also at ]{Dept. of Chemistry, University of Fribourg, 1700 Fribourg, Switzerland}
\email{vweber@t12.lanl.gov}
\author{Anders M. N. Niklasson}
\author{Matt Challacombe}
\affiliation{Los Alamos National Laboratory, Theoretical Division, Los Alamos 87545, New Mexico, USA.}

\date{\today}

\begin{abstract}
A linear scaling method for calculation of the static {\em ab inito} response within self-consistent field theory 
is developed and applied to calculation of the static electric polarizability.  The method is based on 
density matrix perturbation theory [Niklasson and Challacombe, PRL (LJ9425)], obtaining  response functions
directly via a perturbative approach to spectral projection.  The accuracy and efficiency of the linear scaling 
method is demonstrated for a series of three-dimensional 
water clusters at the RHF/6-31G** level of theory.  Locality of the response under a global electric field 
perturbation is numerically demonstrated by approximate exponential decay of derivative density matrix elements.
\end{abstract}


\maketitle

\footnotetext[1]{LA-UR-03-9111: To apear in PRL.}

Linear scaling methods that reduce the computational complexity of 
electronic structure calculations to ${\cal O}(N)$, where $N$ is system size, 
impact disciplines that demand quantum simulation of increasingly large and complex systems
\cite{GGalli96,DBowler97,SGoedecker99,POrdejon00,VGogonea01,SWu02}. 
To date, the most successful and prolific applications of linear scaling technologies are
ground state studies involving empirical model Hamiltonians.  
To a lesser degree, ground state applications using {\em ab initio} models 
with $N$-scaling contribute to a variety of fields, typically at the
Self-Consistent-Field (SCF) level of theory and requiring in some cases parallel 
implementations to reach levels of applicability comparable with empirical models. 
These SCF theories include the Hartree-Fock (HF), Kohn-Sham Density Functional (DF) 
and hybrid HF/DF models.

Beyond ground state methods, little attention has been given 
to linear scaling algorithms for the computation of dynamic and static response 
properties; the latter including the nuclear magnetic shielding tensor \cite{Pulay_1990}, 
the rotational g-tensor \cite{Helgaker_1996}, indirect spin-spin coupling constant 
\cite{Pennington_1991,Malkin_1996}, third order  properties such as the first hyperpolarizability 
\cite{Franky_1997} and polarizability derivatives such as the Raman intensity 
\cite{Lazzeri_2003,Champagne_2001}.  Dynamic  properties may be computed using linear scaling algorithms to propagate the density matrix \cite{SNomura97,CYam03} in 
the time domain, followed by convolution to obtain the spectral response.  In this way, 
Yam, Yokojima and Chen \cite{CYam03} have recently demonstrated linear scaling 
computation of the absorption spectra for one-dimensional polymers at the local 
density level of theory, but requiring $\sim 14,000$ time steps.  In the static zero frequency limit, 
solving the coupled-perturbed self-consistent-field (CPSCF) equations using standard algorithms is likewise difficult for
large systems.
%for systems larger than $\sim 100$ atoms is difficult.  
Several algorithms have therefore been proposed for solving the CPSCF equations 
that may be capable of achieving a reduced scaling better than ${\cal O}(N^3)$ 
\cite{COchsenfeld97,HLarsen01a}.  

In this letter, we develop the density matrix perturbation theory of Niklasson and Challacombe 
\cite{ANiklasson04} for $N$-scaling solution of the {\em ab initio} CPSCF equations, and 
demonstrate the early onset of linear scaling for the accurate calculation of the first electric 
polarizability of three-dimensional systems using large basis sets. This perturbed projection 
method is general and can be extended to a variety of static properties.

Algorithms for linear scaling self-consistent-field (SCF) theory exploit the quantum locality
of non-metallic systems, manifested in the approximate exponential decay of the density matrix 
with atom-atom separation, through the effective use of sparse matrix methods and iterative 
approaches to spectral projection \cite{ANiklasson02A,ANiklasson03}.  This quantum locality should,  in principle, extend also to 
the derivative density matrices central to the CPSCF equations.  Indeed,  
exponential decay of the derivative density matrix within {\em ab initio} SCF theory has been demonstrated 
numerically for local nuclear displacement \cite{Ochsenfeld_1997}. However, standard approaches to the CPSCF equations 
\cite{Pople_1979,Sekino_1986,Dupuis_1991} do not admit exploitation of this locality, as they are based 
on perturbation of the wave function, requiring ${\cal O}(N^3)$ eigensolution and typically ${\cal O}(N^5)$ 
transformation of two-electron integrals into the eigenbasis.   Avoiding both 
eigensolution and integral transformation,  Ochsenfeld and Head-Gordon \cite{Ochsenfeld_1997} and
later Larsen {\em et al.} \cite{Helgaker_2001} proposed iterative solutions to the CPSCF equations 
involving purely non-orthogonal representations.   In both of these approaches, a 
linear system of equations containing commutation relations must be solved.
%This formulation of the CPSCF equations reduces to the Sylvester equation, which using iterative sparse
%methods can be solved in ${\cal O}(N^2)$ \cite{JBrandts01} at best.

Recently, a formulation of density matrix perturbation theory has been proposed 
by Niklasson and Challacombe (NC) \cite{ANiklasson04} that presents a new opportunity for solving 
the CPSCF equations within the context of linear scaling spectral projection \cite{ANiklasson02A,ANiklasson03}.  
The new approach is based on the relationship between the density matrix $\mathcal{D}$ and the effective Hamiltonian 
or Fockian $\mathcal{F}$ through the spectral projector (Heaviside step function) $\mathcal{D}=\theta(\tilde{\mu}I-\mathcal{F})$, 
where the chemical potential $\tilde{\mu}$ determines the occupied states via Aufbau filling.   
Spectral projection can be carried out in a number of ways 
\cite{ANiklasson02A,ANiklasson03,RMcWeeny60,WClinton69,APalser98,GBeylkin99,KNemeth00,AHolas01}, 
with perhaps the best know algorithm being McWeeny's cubic purification scheme \cite{RMcWeeny60}. 
More recently new recursive polynomial expansions of the projector have emerged, 
such as the second order trace correcting (TC2) \cite{ANiklasson02A} and fourth order trace resetting 
(TRS4) \cite{ANiklasson03} purification.  These new methods (TC2 and TRS4) have convergence properties 
that depend only weakly on the band gap, do not require knowledge of the chemical potential
and perform well for all occupation to state ratios. In the NC approach, 
the perturbation expansion is developed within the reference groundstate projector allowing 
order by order collection of terms at each iteration, establishing a quadratically convergent 
sequence for the response functions.  

We now proceed with development of the perturbed projection approach, which will be  
outlined for the concrete case of perturbation by an electric field within the Hartree-Fock (HF) theory.  
However, the perturbed projection method is general and extensible to DF and hybrid HF/DF models, 
and to other static perturbations through high order.

In the following, the indexes $a, b, \dots$ refer to perturbation order, while $i, j, \dots$  mark the iteration count.
The symbols $\mathcal{D},\mathcal{F},\dots$  are  matrices in an orthogonal representation, while
$D,F,\dots$ are the corresponding matrices in a non-orthogonal basis.  The 
transformation between orthogonal and non-orthogonal representations is carried out in ${\cal O}(N)$ using
congruence transformations \cite{JWilkinson65,GStewart73} provided by the AINV algorithm for computing 
sparse approximate inverse Cholesky factors \cite{MBenzi95,MBenzi96,MBenzi01}.  

Within HF theory, the total electronic energy $E_{\rm tot}$ of a molecule in a static electric field $\mathcal{E}$ is
\begin{equation}
   E_{\rm tot}(\mathcal{E})=Tr[D(h^0+\mu \mathcal{E})]
                       +\frac{1}{2}Tr[D(J(D)+K(D))], \label{totalE}
\end{equation}
where $D$ is the density matrix in the electric field, $h^0$ is the core Hamiltonian,  
$\mu$ is the dipole moment matrix, $J(D)$ is the Coulomb matrix and $K(D)$ the exact HF exchange 
matrix.  The total energy may be developed in the perturbation expansion 
\begin{equation}
E_{\rm tot}(\mathcal{E})=E_{\rm tot}(0)+\sum_a\mu_a\mathcal{E}^a+\frac{1}{2}\sum_{ab}\alpha_{ab}\mathcal{E}^a\mathcal{E}^b+\dots,
\end{equation}
 where 
$\alpha_{ab}$ is the first order polarizability, $\mu_a$ is the dipole moment and $\mathcal{E}^a$ is the electric field in
direction $a$.  The polarizability is the second order response of the total energy with respect 
to variation in the electric field \cite{Sekino_1986}
\begin{equation}\label{pol}
   \alpha_{ab}=
   -\frac{\partial^2 E_{\rm tot}}{\partial \mathcal{E}^a\partial \mathcal{E}^b}
   \bigg|_{\mathcal{E}=0}=
   -2Tr[D^a\mu_b].
\end{equation}
The first order density matrix derivative $\mathcal{D}^a$ in the $a$ direction is obtained by variation 
of both the spectral projector 
$\theta$ and the Fockian $\mathcal{F}=\mathcal{F}^{0}+\sum_a\mathcal{E}^{a}\mathcal{F}^{a}+\dots$ as 
 \begin{equation}\label{Step}
   \mathcal{D}^a=\frac{\partial}{\partial \mathcal{E}^a}
   \theta(\tilde{\mu} I-(\mathcal{F}^{0}+\mathcal{E}^{a}\mathcal{F}^{a}))
   \bigg|_{\mathcal{E}=0}.
 \end{equation}

The HF Fockian $F^0$ in the non-orthogonal basis is $F^0=h^0+J(D^0)+K(D^0)$, where the Coulomb matrix $J$ may
be computed in ${\cal O}(N {\rm lg} N )$ with a quantum chemical tree code (QCTC) \cite{MChallacombe97} and the 
exchange matrix $K$ computed in ${\cal O}(N)$ with the ONX algorithm that exploits quantum locality  of $D^0$ \cite{ESchwegler97}.
Likewise the derivative Fockian, $F^a=\mu_a+J(D^a)+K(D^a)$, may be computed with the same algorithms in 
linear scaling time if $D^a$ manifests decay properties similar to $D^0$.  A similar equation holds for the derivative Fockian 
within DF and hybrid HF/DF theories with addition of the exchange-correlation 
matrix $V_{xc}^a(D^0,D^a)$ \cite{Lee_1994}.

In our approach to linear scaling computation of the polarizability $\alpha_{ab}$, the ground state 
density matrix $\mathcal{D}^0$ is computed using a spectral projection algorithm such 
as TC2 \cite{ANiklasson02A} in conjunction with sparse atom-blocked linear algebra \cite{ANiklasson03,MChallacombe00B}.  
Linear scaling is achieved for insulating systems through the dropping (filtering) of atom-atom 
blocks with Frobenious norm below a numerical threshold ($\sim 10^{-4}-10^{-6}$).
At SCF convergence the TC2 algorithm generates a polynomial sequence defining the groundstate projector, 
from which expansion of the derivative density matrix can be obtained term by term.

The derivative density matrix and derivative Fockian depend on each other implicitly and must be 
solved for self-consistently via the coupled-perturbed self-consistent-field (CPSCF) equations.
The necessary and sufficient criteria for convergence of the CPSCF equations are 
$[\mathcal{F}^{a},\mathcal{D}^{0}]+[\mathcal{F}^{0},\mathcal{D}^{a}]=0$ and 
$\mathcal{D}^{a}=\mathcal{D}^{a} \mathcal{D}^{0}+\mathcal{D}^{0} \mathcal{D}^{a}$ \cite{Furche_2001}.
Solution of the CPSCF equations with perturbed projection involves the steps
\begin{subequations}
\begin{eqnarray}
&&     F^a_{n}=\mu_a+J(D^a_n)+K(D^a_n) \label{FockBuild} \\
&&     \displaystyle\widetilde{F}^a_{n}=\sum_{k=n-m}^{n}c_k F^a_{k} \label{DDIIS} \\
&&     \displaystyle\mathcal{D}^a_{n+1}=\frac{\partial}{\partial \mathcal{E}^a}
     \theta(\tilde{\mu}I-(\mathcal{F}^{0}
     +\mathcal{E}^{a}\widetilde{\mathcal{F}}^{a}_n))
     \bigg|_{\mathcal{E}=0} \label{DDeriv}
   \end{eqnarray} 
\end{subequations}
with starting point $D^a_0=0$. In step~(\ref{FockBuild}),  $F^a_n$ is constructed in 
${\cal O}(N)$ using the QCTC \cite{MChallacombe97} and ONX \cite{ESchwegler97} algorithms in 
{\sc MondoSCF} \cite{MondoSCF}.  Next, 
Weber and Daul's DDIIS algorithm for convergence acceleration of the CPSCF equation
\cite{Weber_2003} is used to optimize the $c_k$ coefficients in step~(\ref{DDIIS}), keeping the last $m$
steps in the extrapolation. 
Then, the density matrix derivative $\mathcal{D}^a_{n+1}$ is obtained in step~(\ref{DDeriv}) as 
$\mathcal{D}^a_{n+1}=\lim_{i\to\infty}\mathcal{X}^a_{i}$ via
the NC density matrix perturbation theory,
based on the TC2 projector:
\begin{equation}\label{PP1}
\left.
\begin{array}{ll}
\mathcal{X}^a_{i+1}&=\mathcal{X}^a_{i}\mathcal{X}^0_{i}+\mathcal{X}^0_{i}\mathcal{X}^a_{i}\\
\mathcal{X}^0_{i+1}&=(\mathcal{X}^0_{i})^2
\end{array} 
\right\}\quad {\rm Tr}[\mathcal{X}^0_{i}]\ge N_e 
\end{equation}
or 
\begin{equation}\label{PP2}
\left.
\begin{array}{ll}
\mathcal{X}^a_{i+1}&=2\mathcal{X}^a_{i}-\mathcal{X}^a_{i}\mathcal{X}^0_{i}-\mathcal{X}^0_{i}\mathcal{X}^a_{i} \\
\mathcal{X}^0_{i+1}&=2\mathcal{X}^0_{i}-(\mathcal{X}^0_{i})^2
\end{array} 
\right\}\quad {\rm Tr}[\mathcal{X}^0_{i}]< N_e.
\end{equation}
The matrices initiating the sequence are obtained from $\mathcal{F}^0$
and  $\mathcal{F}^a$ by appropriately 
compressing their spectrum into the domain of convergence \cite{ANiklasson02A} using
\begin{equation}
\mathcal{X}^0_{0}=\frac{\mathcal{F}_{max}-\mathcal{F}^0}{\mathcal{F}_{max}-\mathcal{F}_{min}}~{\rm and}~
\mathcal{X}^a_{0}=\frac{\mathcal{F}^a_{n}}{\mathcal{F}_{min}-\mathcal{F}_{max}},
\end{equation}
where $\mathcal{F}_{min}$ and $\mathcal{F}_{max}$ are approximate upper and lower bounds to the eigenvalues of $\mathcal{F}^0$.

Recursion of the perturbed projection sequence is stopped when the change 
$\left| \mathcal{X}^a_{i+1}-\mathcal{X}^a_i \right|$ becomes small. Having solved for $D^a_n$, 
the next Fock matrix derivative $F^a_{n+1}$ is built, and the iteration continues until 
self-consistency, when the density matrix derivative $D^a_n$ and the desired property 
(e.g. the polarizability $\alpha_{ab}=-2Tr[D^a_n\mu_b]$) have reached a sufficient level of accuracy.

\commentoutA{
\begin{figure}[t]
\caption{\protect  Total CPU time of the fifth CPSCF iteration for the water cluster sequence with 
         the 6-31G and 6-31G** basis sets and the {\tt GOOD} and {\tt TIGHT} numerical thresholds (see text) 
         controlling numerical precision of the result.  The lines are fits to the last three and four points, respectively.}\label{scaling}
\resizebox*{3.5in}{!}{\includegraphics[angle=-90.00]{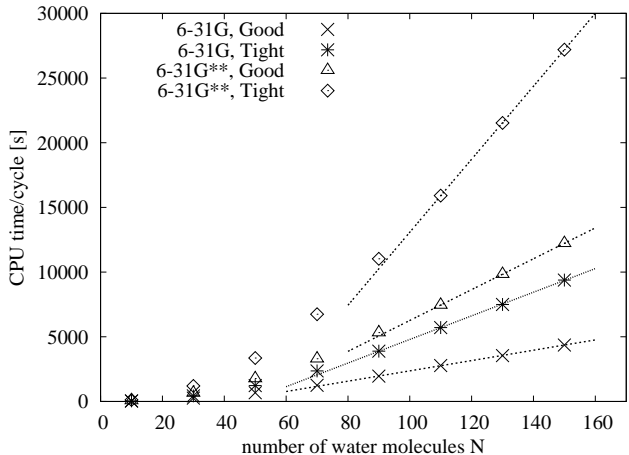}}
\end{figure}
}

We have implemented these methods in the {\sc MondoSCF} suite of linear scaling quantum chemistry programs \cite{MondoSCF},
and performed polarizability calculations on a series of water clusters up to (H$_2$O)$_{150}$.  These clusters were
obtained by carving a spherical region out of a snapshot from a periodic classical  molecular dynamics simulation of water 
at standard liquid density, and have been used previously in a number of scaling tests 
\cite{ANiklasson03,MChallacombe97,ESchwegler97}.

Calculations have been carried out at both the RHF/6-31G and RHF/6-31G** levels of 
theory and with both the {\tt GOOD} and {\tt TIGHT} thresholding parameter sets that control precision of the 
linear scaling algorithms, corresponding to matrix thresholds of $10^{-5}$ and $10^{-6}$, respectively.  
These calculations were carried out on a single Intel Xeon 2.4GHz processor running RedHat Linux 8.0 and  executables compiled 
with Portland Group Fortran Compiler pgf90 4.0-2 \cite{PGF90}. In Fig.~\ref{scaling}, the 
total CPU time for the fifth CPSCF cycle (including build time for $\mathcal{F}^a$, 
iterative construction of $\mathcal{D}^a$ and all intermediate 
steps including congruence transformation) is shown for the RHF/6-31G and RHF/6-31G** series 
of water clusters.  Convergence of the CPSCF equations for these systems are typically 
achieved in about 10 cycles, independent of cluster size, basis set or matrix threshold.  
In Table \ref{tab:Polari_Values},  the corresponding average water cluster polarizabilities computed with the  {\sc MondoSCF}
algorithms are listed and compared to the those obtained with the {\sc GAMESS} quantum chemistry package \cite{gamess} 
at the RHF/6-31G level of theory.  Figure \ref{fig:DPrimeZ_150_6-31G} shows the magnitude of density matrix 
derivative atom-atom blocks as a function of atom-atom distance under global perturbation by a static electric field.

\commentoutA{
\begin{table}[t]
\caption{\protect Average polarizabilities $\bar{\alpha}=(\alpha_{xx}+\alpha_{yy}+\alpha_{zz})/3N_{H_20}$
         in a.u.~for a sequence of water clusters at the RHF/6-31G and RHF/6-31G** levels of theory.
         A comparison is made between results obtained with the {\sc GAMESS} quantum chemistry program
         \cite{gamess} and those calculated with {\sc MondoSCF} using different numerical approximations
         (see text) controlling precision of the linear scaling algorithms.}\label{tab:Polari_Values}
\begin{tabular}{cccccc}
\toprule 
      &\multicolumn{1}{c}{6-31G\footnote[1]{\sc GAMESS}}
      &\multicolumn{2}{c}{6-31G\footnote[2]{{\sc MondoSCF}}}
      &\multicolumn{2}{c}{6-31G**$^b$}\\
      $N_{H_20}$ &          & {\tt GOOD}     & {\tt TIGHT}    &  {\tt GOOD}    & {\tt TIGHT}   \\
      \hline
      10  & 4.569083 & 4.569918 & 4.569102 & 5.479161 & 5.479049  \\
      30  & 4.673213 & 4.673208 & 4.673227 & 5.585293 & 5.585280  \\
      50  & 4.703540 & 4.703512 & 4.703568 & 5.623057 & 5.622830  \\
      70  & $-$      & 4.732207 & 4.732279 & 5.654646 & 5.654747  \\
      90  & $-$      & 4.775002 & 4.775024 & 5.695435 & 5.695564  \\
      110 & $-$      & 4.780718 & 4.780809 & 5.698338 & 5.698447  \\
      130 & $-$      & 4.786383 & 4.786437 & 5.704859 & 5.704947  \\
      150 & $-$      & 4.775124 & 4.775231 & 5.693268 & 5.693447  \\
\botrule
\end{tabular}
\end{table}

\begin{figure}
  \caption{\protect
    Magnitudes of atom-atom blocks of the RHF/6-31G density matrix derivative
    in the z direction with the separation of basis function centers for $(H_2O)_{150}$.
    The density matrix derivative has been converged using a {\tt TIGHT} accuracy level (e.g. 
    a matrix threshold of $10^{-6}$ au).
  }\label{fig:DPrimeZ_150_6-31G}
\resizebox*{3.5in}{!}{\includegraphics[angle=-90.00]{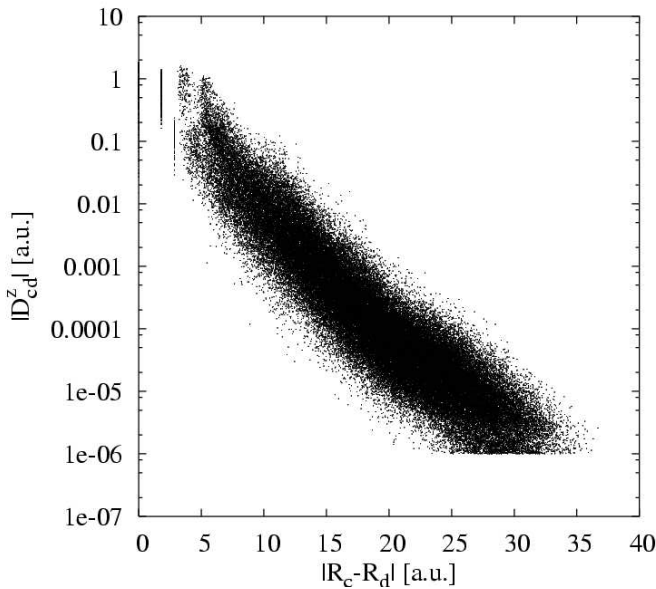}}
\end{figure}
}

These results demonstrate an onset of linear scaling as early as 70 water molecules
for properties with 4 digits of precision (RHF/6-31G at {\tt GOOD}). 
While the CPSCF equations must be solved iteratively with perturbed projection,  
the number of CPSCF cycles is $\sim 10$ when using DDIIS acceleration on well behaved systems.
Using an incomplete sparse linear algebra with thresholding is a particular advantage of the 
present implementation, as the small gap limit correctly leads to ${\cal O}(N^3)$ 
algorithms while preserving accuracy.  In contrast, methods that employ radial cutoffs
incorrectly retain $N$-scaling in this limit at the sacrifice of accuracy. 
However, the iterative approach proposed here for solution of the CPSCF equations is prone to the same 
instabilities encountered by the SCF equations in the small gap limit.  

We have presented a simple and efficient algorithm for solution of the coupled-perturbed 
self-consistent-field (CPSCF) equations in the context of spectral projection and the static 
electric polarizability. Unique features of the perturbed projection algorithm include linear 
scaling, simplicity, numerical stability and quadratic convergence in computation of the derivative 
density matrix.  

We have shown that the density matrix response is local upon {\em global} electric perturbation, 
corresponding to an approximate exponential decay of matrix elements. A similar exponential decay 
in the first order response corresponding to the {\em local} nuclear displacement has previously been
demonstrated by Ochsenfeld and Head-Gordon \cite{Ochsenfeld_1997}.   These key observations are expected to
hold generally for both local and global perturbations to insulating systems.  The implication of these 
results are that the perturbed projection algorithm described in steps (\ref{FockBuild}-\ref{DDeriv}) and
Eqs.~(\ref{PP1}-\ref{PP2}) can be easily extended to the linear scaling computation of higher order 
response functions, DF and HF/DF models and a large class of static molecular properties such as the 
nuclear magnetic shielding tensor (NMR shift), indirect spin-spin coupling and the electronic g-tensor.
We note also that the method is not unique to the TC2 generator or {\sc MondoSCF} $N$-scaling algorithms, 
but can be straightforwardly extended to other purification schemes such as TRS4 \cite{ANiklasson03} as
well as other electronic structure programs.

This work has been supported by the US Department of Energy 
under contract W-7405-ENG-36 and the ASCI project.  
The Advanced Computing Laboratory of Los 
Alamos National Laboratory is acknowledged.
All the numerical computations have been
performed on computing resources located at this facility.

\bibliography{perturbedprojector}

\end{document}